\documentclass[twocolumn,aps,prb,showpacs,amsmath,amssymb,floatfix,superscriptaddress]{revtex4-1}

\usepackage{amsmath}
\usepackage{amssymb}
\usepackage{float}
\usepackage{bbold} 
\usepackage{url}
\usepackage{graphicx}

\begin{document}
\title{Photonic lattices of coaxial cables: flat bands and artificial magnetic fields}

\author{Christopher Oliver}
\affiliation{School of Physics and Astronomy, University of Birmingham, Edgbaston, Birmingham, UK, B15 2TT}
\author{Denis Nabari}
\affiliation{Dipartimento di Fisica, Università di Trento I-38123 Trento, Italy}
\author{Hannah M. Price}
\affiliation{School of Physics and Astronomy, University of Birmingham, Edgbaston, Birmingham, UK, B15 2TT}
\author{Leonardo Ricci}
\affiliation{Dipartimento di Fisica, Università di Trento I-38123 Trento, Italy}
\author{Iacopo Carusotto}
\affiliation{Pitaevskii BEC Center, INO-CNR and Dipartimento di Fisica, Università di Trento I-38123 Trento, Italy}

\begin{abstract}
   We propose the use of networks of standard, commercially-available coaxial cables as a platform to realize photonic lattice models. As a specific example, we consider a brick wall lattice formed from coaxial cables and T-shaped connectors. We calculate the dispersion of photonic Bloch waves in the lattice: we find a repeated family of three bands, which include a flat band and two Dirac points. We then demonstrate a method to displace the Dirac points, leading to an induced artificial gauge field, and a method to energetically isolate the flat band. Our results readily suggest that the interplay of nonlinearities and non-trivial topology are a natural avenue to explore in order to unlock the full power of this proposed platform.
\end{abstract}
\date{\today}

\maketitle

\section{Introduction}

In recent years, circuit QED has emerged as a leading approach to both quantum simulation and quantum computing~\cite{Blais2020}. In the quantum simulation context, the important degrees of freedom are the quantized microwave photons and the most common experimental implementation is based on co-planar waveguide (CPW) resonators fabricated on the surface of a solid-state substrate, which can be thought of as an integrated equivalent of a coaxial cable. CPW segments are then coupled together via suitable elements such as capacitors, leading to an effective tight-binding model for microwave photons. Strong photon-photon interactions can then be mediated via superconducting Josephson tunnel junction elements. The above framework can then be deployed to create `photonic materials', in which the microwave photons hop as in a tight-binding model and strongly interact with each other in order to realise some chosen target quantum many-body Hamiltonian~\cite{Carusotto2020}.

The crucial feature of such photonic materials is that, in stark contrast to electrons in a tight-binding solid, the properties of the lattice do not depend on the precise geometry of the system but only on its connectivity. As such, the CPW segments can be arbitrarily bent to fit on the sample surface. This property allows for the engineering of exotic models with complicated connectivities and/or non-Euclidean effective geometries by simply changing the wiring of the circuit~\cite{Kollar2019}. This feature enables control over both the local geometry and the global topology, for example, by introducing periodic boundary conditions~\cite{Ningyuan2015}. A famous example is the realisation of models in which the particles effectively live in hyperbolic spaces, allowing insight into quantum field theories on curved spaces~\cite{Kollar2019, Boettcher2020, Bienias2022}. In some of these geometries, the particles also experience destructive interference in the hopping processes, leading to the emergence of flat bands in which interaction effects dominate.

However, CPW resonator lattices suffer from some practical drawbacks. In particular, the circuits are fabricated using a relatively complicated lithographic procedure, which precludes any dynamical changes to the lattice geometry on-the-fly, and they need to be cooled to superconducting temperatures before use. Given that CPW resonators are integrated analogs of coaxial cables, a natural question to ask is whether we could investigate similar physics by constructing lattices formed by standard, commercially-available coaxial cables connected by standard connector elements. Such a system could have the major advantage of the CPW lattices, namely the decoupling of the relevant physical parameters from the circuit geometry, without the practical limitations listed above. 

In this work, we demonstrate the potential of lattices of coaxial cables coupled by T-shaped connector elements in simulating diverse physical phenomena by modelling a brick wall lattice. As was noticed in Ref.~\onlinecite{Kollar2019}, the effective photonic lattice is the dual of the physical lattice shown in Fig.~\ref{fig.1}: the sites are located at the coaxial cables and each T-shaped connector element provides hopping between the coaxial cables impinging on it. The effective lattice then has a Kagome geometry. 
Our work builds upon recent results addressing the realisation of the SSH model in a simplified 1D version of the coaxial cable lattice~\cite{Nabari2021,Balduzzi2022}. A related investigation was reported in Ref.~\onlinecite{Whittaker2021}. 

Our investigations complement existing studies that use lumped-element electric circuits to simulate lattice models. Here,  lattices are formed from a network of inductors, resistors and capacitors (LRC circuits), where the circuit Laplacian plays the role of a Hamiltonian for a lattice model~\cite{Lee2018}. Such lattices have been used to realise celebrated topological models such as the SSH model~\cite{Su1979, Lee2018} and models in three~\cite{Lu2019} or even higher dimensionality~\cite{Price2020, Yu2020, Wang2020}. We also note that there are a small number of works from around 20 years ago that consider the properties of coaxial cable lattices~\cite{Dobrzynski1998, Poirier2001, Schneider2001, Hache2002, Hache2004}. These works consider, for instance, the effect of nonlinearities and defects in lattices of this type, but they are more focused on applications in photonics, as opposed to the simulation of topological lattice models. Finally, topological photonic models based on network configurations similar to ours have been considered in the optical domain in Ref.~\onlinecite{shi2017topological}.

This work is structured as follows. We begin by considering the simplest case in which all the cables are identical (the `uniform' case), and calculate the dispersion of Bloch waves in the lattice. We find two main features of interest: Dirac points formed by the dispersive bands, and flat bands that are not energetically isolated. We then study two different methods to control these features. We firstly show that the location and gap of the Dirac points can be controlled by tuning the impedance of the cables, suggesting connections to the theory of artificial magnetic fields in strained graphene. Secondly, we show that the flat bands can be energetically isolated by controlling the cable lengths, suggesting that the inclusion of nonlinear circuit elements is a promising route to explore in the future. Conclusions are finally drawn.

\section{A brick wall lattice}
\label{sec:uniform_mt}
\begin{figure}
    \includegraphics[scale=0.25]{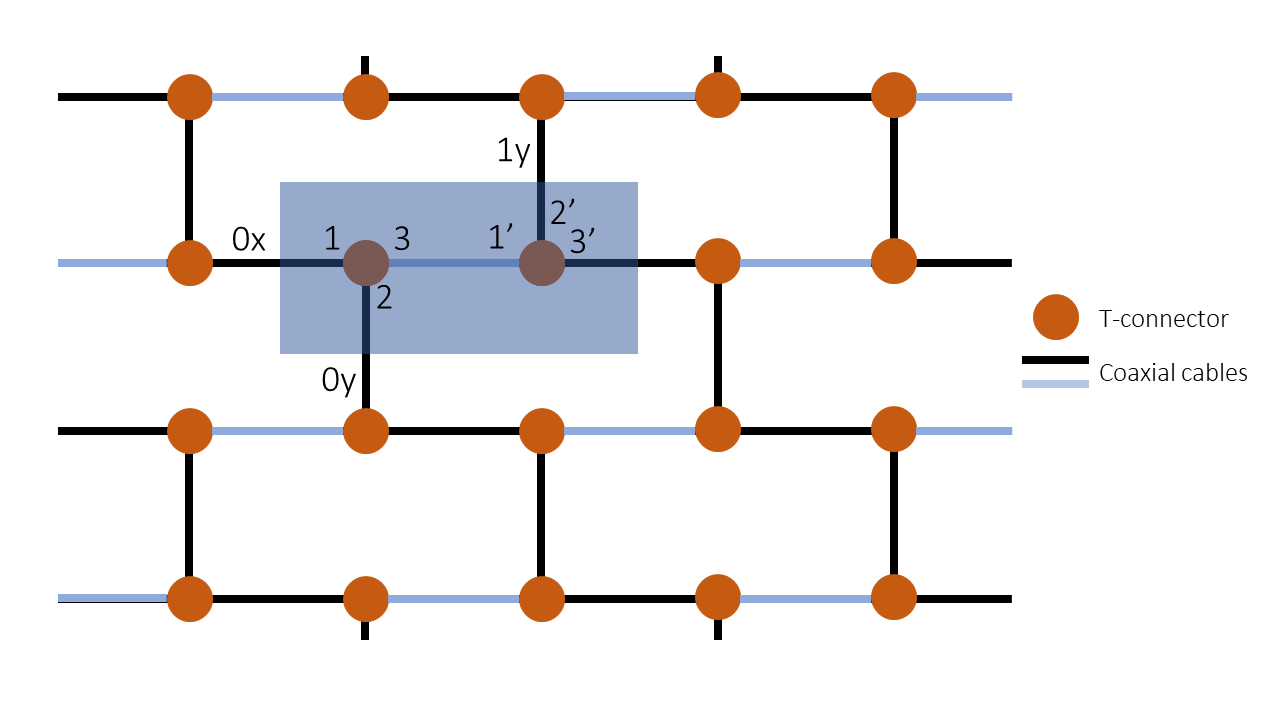}
    \caption{Sketch of the physical network under investigation. The coaxial cables (light blue and black) form the edges of a brick wall lattice, and the T-shaped connector elements (orange) form the sites. The unit cell chosen is shown by the blue rectangle, with the labels of various modes shown. In some later calculations, the properties of the cable running between the two connectors in each unit cell are changed relative to the others, so these are highlighted in light blue.}
    \label{fig.1}
\end{figure}
We begin by considering a brick wall lattice formed from coaxial cables and T-shaped connector elements, in which the connectors play the role of the brick wall lattice sites, and the edges between the sites are the cables. In this first instance, the cables all have identical length $l$ and impedance $Z$. This situation is shown in Fig.~\ref{fig.1}. 

We calculate the dispersion of the photonic Bloch waves in three main steps:
\begin{enumerate}
    \item Derive a $4 \times 4$ scattering matrix connecting the in- and out-going modes for a single unit cell of the brick wall lattice (blue rectangle in Fig.~\ref{fig.1}), by considering the phase picked up by a wave of frequency $\omega$ as it propagates through the cable segments, as well as the scattering at the connector elements.
    \item Impose periodic boundary conditions on the unit cell and introduce Bloch momenta $k_x$ and $k_y$.
    \item Search for non-trivial solutions of the resulting matrix equation, leading to an equation that can be solved to obtain the dispersion $\omega(k_x, k_y)$.
\end{enumerate}
\begin{figure*}
    \includegraphics[scale=0.5]{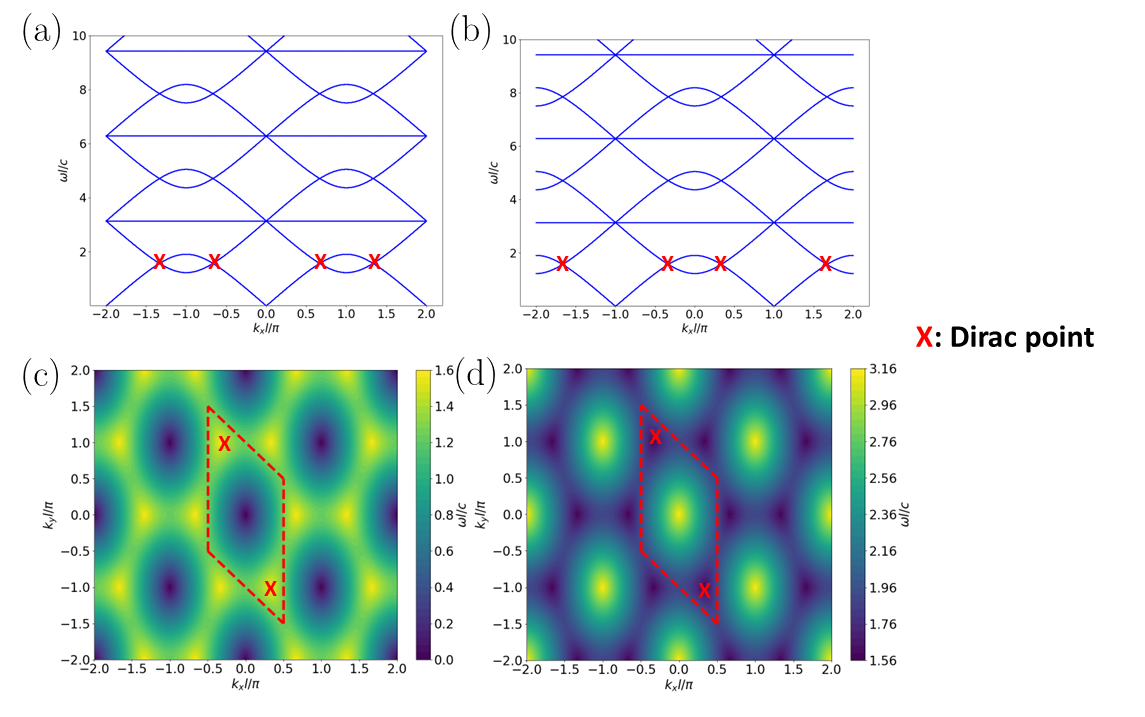}
    \caption{Band structure for the brick wall lattice with all cables identical. (a) and (b): Cuts of the bands at $k_yl = 0$ and $k_yl = \pi$ respectively, showing the flat bands at $\omega l/c = n\pi$ ($n = 0,1,2,...$) and the two repeated dispersive bands. (c) and (d): Contour plots of the two lowest dispersive bands (located in the interval $\omega l/c \in [0,\pi]$), with nearly-free-photon-like linear dispersions at low frequencies and two gapless Dirac points (red) with momenta $(\pm \pi / 3, \mp\pi)$. The red dashed line indicates the first Brillouin zone.}
    \label{fig.2}
\end{figure*}
The essential ingredients in Step 1 are the equations describing the propagation of waves through the cable segments, and their scattering at the connector elements. Working from left-to-right and from bottom-to-top in the unit cell in Fig.~\ref{fig.1}, we have:
\begin{equation}
    \begin{pmatrix}
        I_1 \\
        O_1
    \end{pmatrix}
    = 
    U_{l/2} 
    \begin{pmatrix}
        I_{0x} \\
        O_{0x}
    \end{pmatrix}
    \equiv
    \begin{pmatrix}
        e^{i\frac{\omega}{c}\frac{l}{2}} & 0 \\
        0 & e^{-i\frac{\omega}{c}\frac{l}{2}}
    \end{pmatrix}
    \begin{pmatrix}
        I_{0x} \\
        O_{0x}
    \end{pmatrix},
\end{equation}
and:
\begin{equation}
    \begin{pmatrix}
        I_2 \\
        O_2
    \end{pmatrix}
    = 
    U_{l/2} 
    \begin{pmatrix}
        I_{0y} \\
        O_{0y}
    \end{pmatrix}
\end{equation}
describing the propagation of the modes at the left-hand edges towards the connector. These modes then scatter at the connector, which is described by the scattering matrix $S$:
\begin{equation}
    \begin{pmatrix}
        I_1 \\
        I_2 \\
        I_3 
    \end{pmatrix}
    = S
    \begin{pmatrix}
        O_1 \\
        O_2 \\
        O_3 
    \end{pmatrix}=
    \frac{1}{3}
    \begin{pmatrix}
        -1 & 2 & 2 \\
        2 & -1 & 2 \\
        2 & 2 & -1
    \end{pmatrix} 
    \begin{pmatrix}
        O_1 \\
        O_2 \\
        O_3 
    \end{pmatrix}\,.
    \label{eqn:left_T}
\end{equation}
where we have assumed that all the cable impedances are identical and the T-shaped connectors are symmetric.
. 
We then describe the propagation of the modes along the central cable:
\begin{equation}
    \begin{pmatrix}
        I_{1^{\prime}} \\
        O_{1^{\prime}}
    \end{pmatrix}
    = 
    U_l 
    \begin{pmatrix}
        O_3 \\
        I_3
    \end{pmatrix}.
\end{equation}
We also have the same scattering matrix for the right-hand connector:
\begin{equation}
    \begin{pmatrix}
        I_{1^{\prime}} \\
        I_{2^{\prime}} \\
        I_{3^{\prime}} 
    \end{pmatrix}
    = S
    \begin{pmatrix}
        O_{1^{\prime}} \\
        O_{2^{\prime}} \\
        O_{3^{\prime}} 
    \end{pmatrix},
\end{equation}
and, finally, we relate the modes at the right-hand unit cell edge to these scattered modes:
\begin{equation}
    \begin{pmatrix}
        I_{1y} \\
        O_{1y}
    \end{pmatrix}
    = 
    U_{l/2}^{-1} 
    \begin{pmatrix}
        I_{2^{\prime}} \\
        O_{2^{\prime}}
    \end{pmatrix}
\end{equation}
and:
\begin{equation}
    \begin{pmatrix}
        I_{1x} \\
        O_{1x}
    \end{pmatrix}
    = 
    U_{l/2}^{-1} 
    \begin{pmatrix}
        I_{3^{\prime}} \\
        O_{3^{\prime}}
    \end{pmatrix}
\end{equation}
We then combine these various equations to calculate an effective scattering matrix $\Sigma$ for the entire unit cell:
\begin{equation}
    \begin{pmatrix}
        I_{0x} \\
        I_{0y} \\
        I_{1x} \\
        I_{1y}
    \end{pmatrix}
    = \Sigma
    \begin{pmatrix}
        O_{0x} \\
        O_{0y} \\
        O_{1x} \\
        O_{1y} 
    \end{pmatrix} = \begin{pmatrix}
        \Sigma_1 & \Sigma_2 & \Sigma_3 & \Sigma_3 \\
        \Sigma_2 & \Sigma_1 & \Sigma_3 & \Sigma_3 \\
        \Sigma_3 & \Sigma_3 & \Sigma_1 & \Sigma_2 \\
        \Sigma_3 & \Sigma_3 & \Sigma_2 & \Sigma_1
    \end{pmatrix}
    \begin{pmatrix}
        O_{0x} \\
        O_{0y} \\
        O_{1x} \\
        O_{1y} 
    \end{pmatrix}
    \label{eq:sigma}
\end{equation}
with:
\begin{eqnarray}
    \Sigma_3 &\equiv& \frac{4}{3}\frac{e^{-i\frac{\omega}{c}l}}{3e^{i\frac{\omega}{c}l} - \frac{1}{3}e^{-i\frac{\omega}{c}l}}, \\ \Sigma_1 &\equiv& \frac{1}{3}e^{-i\frac{\omega}{c}l}(-1 - \Sigma_3) \\ \Sigma_2 &\equiv& \frac{1}{3} e^{-i\frac{\omega}{c}l}(2 - \Sigma_3).
    \end{eqnarray}

Now that the propagation of waves with frequency $\omega$ within a single unit is described, we can proceed with Step 2, which consists of introducing the lattice into the problem and imposing the corresponding periodic boundary conditions:
\begin{equation}
   \begin{pmatrix}
       I_{1x} \\
       O_{1x}
   \end{pmatrix} 
   =
   e^{2ik_xl}
   \begin{pmatrix}
       O_{0x} \\
       I_{0x}
   \end{pmatrix} \textrm{~and~}   
   \begin{pmatrix}
       I_{1y} \\
       O_{1y}
   \end{pmatrix} 
   =
   e^{i(k_x + k_y)l}
   \begin{pmatrix}
       O_{0y} \\
       I_{0y}
   \end{pmatrix},
\end{equation}
where we have introduced the Bloch momenta $k_x$ and $k_y$. Combining these boundary conditions with the $4 \times 4$ scattering matrix and requiring non-trivial solutions to the resulting $4 \times 4$ matrix equation leads to the equation:
\begin{equation}
    \det\left({P_{BC}^{-1}\Sigma P_{BC}
    \begin{pmatrix}
        0 & \mathbb{1}_2 \\
        \mathbb{1}_2 & 0
    \end{pmatrix}
    - \mathbb{1}_4}\right)
     = 0,
     \label{eq:main}
\end{equation}
where $\mathbb{1}_n$ is the $n \times n$ identity matrix and $P_{BC} \equiv \text{diag}(1,1,\exp(2ik_xl),\exp(i(k_x + k_y)l)$. 

As a Step 3, Eq.~\ref{eq:main} can then be solved analytically to obtain the dispersion $\omega(k_x, k_y)$.

As is shown in Fig.~\ref{fig.2}, we find a set of three principal bands $\omega_{PV}(k_x, k_y)$ in the interval $\omega l/c \in [0, \pi]$. As is typical of Kagome lattices, these include a flat band with $\omega l/c = \pi$ and two dispersive bands, whose dispersions are given by:
\begin{equation}
    \cos(2\omega l/c) = \frac{1}{9}(-3 + 4(\cos(2k_xl) + 2\cos(k_xl)\cos(k_yl))).
    \label{eqn:dispersive}
\end{equation}
Like in the optical model of Ref.\onlinecite{shi2017topological}, the three principal bands are then periodically repeated in $\omega$, with $\omega(k_x, k_y)l/c = \omega_{PV}(k_x, k_y)l/c + n\pi$, for $n = 0,1,2,...$.%
\begin{figure*}[t]
    \includegraphics[scale=0.5]{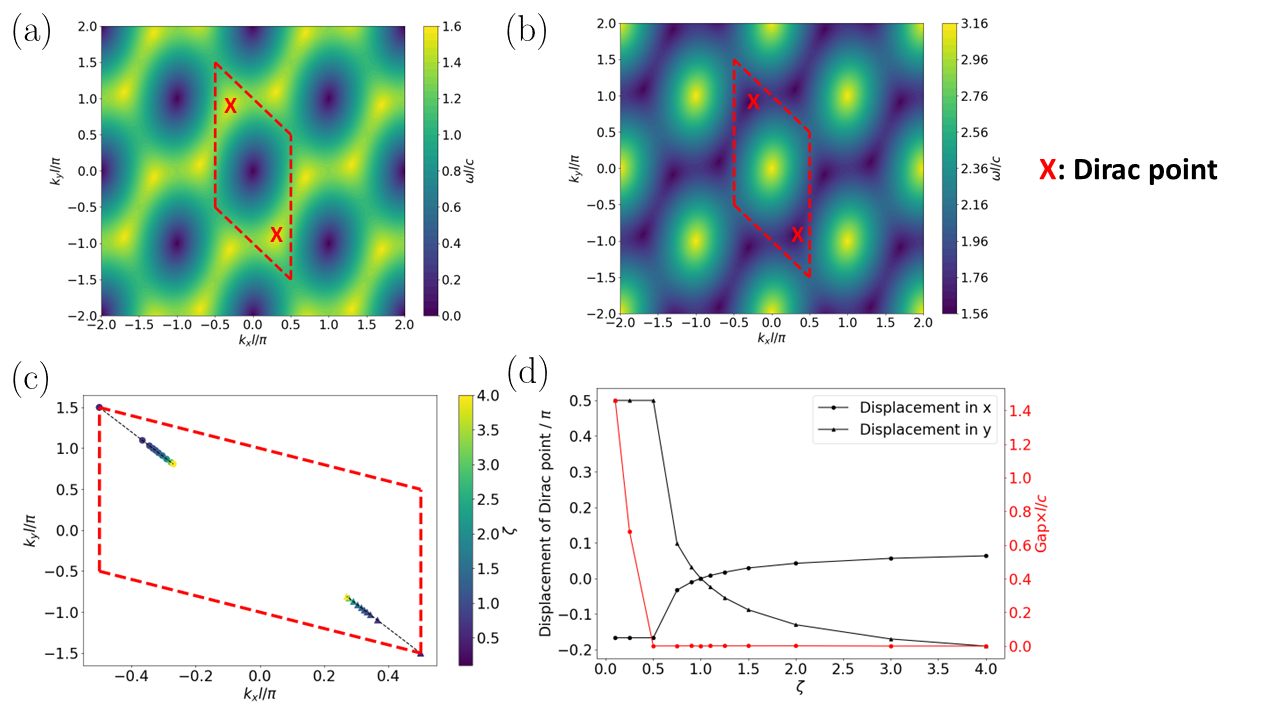}
    \caption{Results for the brick wall lattice with the central cable impedance detuned in each unit cell. (a) and (b): contour plots of the two dispersive bands for the case where $\zeta = Z_2/Z_1= 3/2$, with nearly-free-photon-like linear dispersions close to the origin and the two Dirac points (red) in each Brillouin zone displaced by the detuning. The flat bands still exist at the same frequency as the uniform case and remain gapless at the origin. (c): trajectories taken by the two principal Dirac points (circles and triangles) around the first Brillouin zone (red dashed line) as $\zeta$ is varied. The black dashed line is a guide to the eye. As $\zeta$ is decreased below some critical value, the Dirac points merge with those in neighbouring Brillouin zones and annihilate in pairs at the points $(\pm \pi / 2, \mp 3\pi / 2)$, opening a gap between the dispersive bands. (d): $k_x$ and $k_y$ displacements of the top-left Dirac point from its position when $\zeta = 1$ as $\zeta$ varies (black curves, left axis). Also shown is the gap between the Dirac points as a function of $\zeta$ (red curve, right axis), showing a gap opening as the Dirac points annihilate. After the Dirac points merge, the position plotted is defined by the maximum of the lower band/minimum of the upper band.}
    \label{fig.3}
\end{figure*}
A straightforward Taylor expansion of the two dispersive bands around $k_x = k_y = 0$ shows that the dispersion is linear there, indicating that we are working in the nearly-free photon limit. More precisely, we write $k_xl = \varepsilon_x \ll 1$ and $k_yl = \varepsilon_y \ll 1$. We then Taylor-expand the right-hand side of Eq.~\ref{eqn:dispersive}, and then expand the left-hand side about $\omega = 0$ for the lower band and about $\omega l/c = \pi$ for the upper band. This procedure produces
\begin{equation}
    \omega l/c = \sqrt{\frac{2}{3}\varepsilon_x^2 + \frac{2}{9}\varepsilon_y^2} \textrm{~and~}     \pi - \sqrt{\frac{2}{3}\varepsilon_x^2 + \frac{2}{9}\varepsilon_y^2}. 
\end{equation}
for the lower and the upper band, respectively.

But the two main features of interest in our dispersion are the following. Firstly, the two dispersive bands meet each other at two gapless Dirac points in each Brillouin zone (red crosses in Fig.~\ref{fig.2}), with the momenta $(\pm\pi / 3, \mp\pi)$ in the first Brillouin zone (red dashed line). A Taylor expansion shows that the frequency varies linearly with momentum in the vicinity of these Dirac points. Secondly, the flat band meets the two dispersive bands at the origin (Fig.~\ref{fig.2}(a)). As we will soon see, the experimental control of these features would unlock a wide variety of physics that could be engineered in this platform. 

\section{Displacing the Dirac points}
\label{sec:imp}
\begin{figure*}[t]
    \includegraphics[scale=0.55]{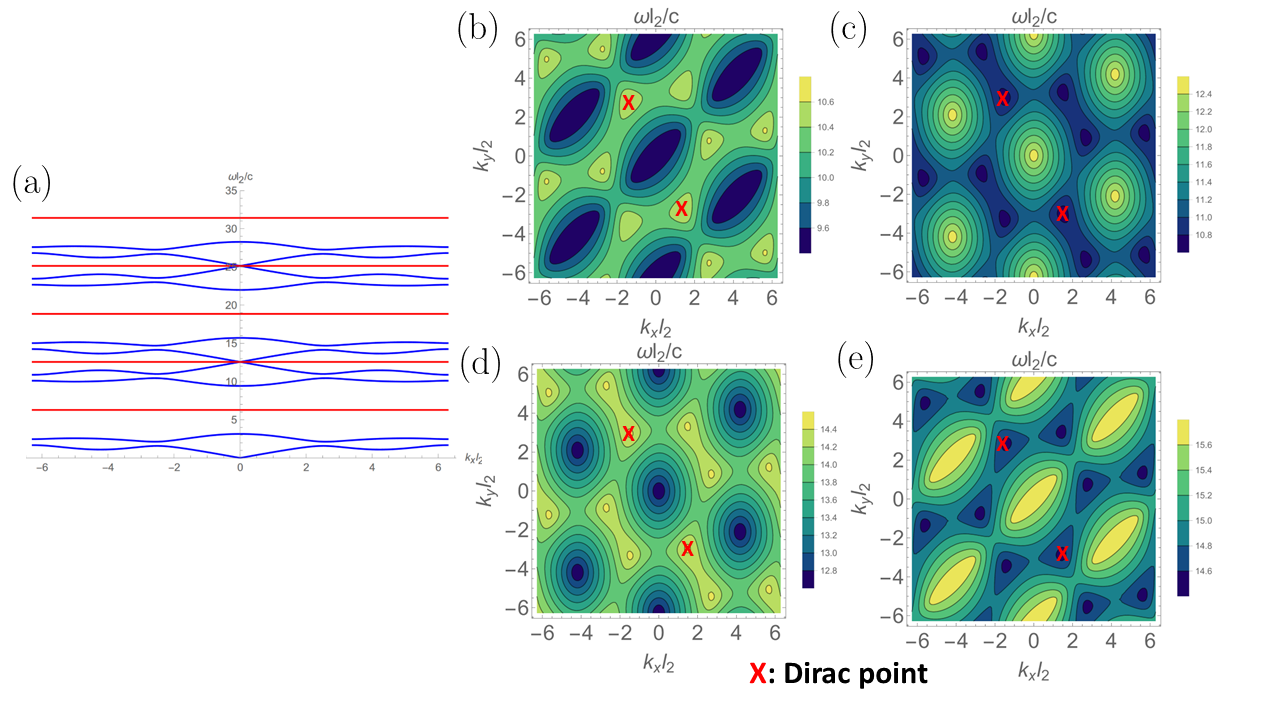}
    \caption{(a): Cut of the $\lambda = 2$ length-detuned brick wall lattice dispersion through $k_y = 0$, showing the first few periods of the bands. We see a family of dispersive bands that is periodic in $\omega$ (blue) and a family of flat bands interleaved with these (red). A subset of these flat bands are energetically isolated ($\omega l_2/c = 2\pi, 6\pi, 10\pi, ...$, i.e. odd multiples of $2\pi$), which can be seen by checking the whole band structure. (b) - (e): Contour plots of the four dispersive bands. More precisely, here we show the four bands in the $n = 1$ sector, with $\omega l_2/c = \omega_{PV} l_2/c + 4\pi n$. The panels are ordered in increasing $\omega$, so panel (a) corresponds to the bottom $n = 1$ band, etc. Dirac points in the first Brillouin zone are marked with a red cross.}
    \label{fig.4}
\end{figure*}
We consider changing the impedance of the `central' cable running between the two connectors in each unit cell, such that the other cables have impedance $Z_1$ and the central cable has impedance $Z_2$ (Fig.~\ref{fig.1}). The calculation we carry out to find the Bloch wave dispersion is generally the same as in the uniform case, but the different impedances lead to different reflection and transmission coefficients between different ports of the T-connectors. 
We now have the T-connector scattering matrix:
\begin{equation}
    \bar{S} = 
    \begin{pmatrix}
        r & t^{\prime} & t \\
        t^{\prime} & r & t \\
        t & t & r^{\prime}
    \end{pmatrix},
\end{equation}
where:
\begin{align}
    r = -\frac{1}{1 + 2\zeta},\\
    r^{\prime} = \frac{1 - 2\zeta}{1 + 2\zeta},\\
    t^2 = \frac{1}{2}(1 - {r^{\prime}}^2),\\
    {t^{\prime}}^2 = 1 - r^2 - t^2,
\end{align}
in terms of the ratio $\zeta \equiv Z_2/Z_1$ of the coaxial cable impedances. We then have:
\begin{equation}
    \begin{pmatrix}
        I_1 \\
        I_2 \\
        I_3
    \end{pmatrix}
    = \bar{S}
    \begin{pmatrix}
        O_1 \\
        O_2 \\
        O_3
    \end{pmatrix} \textrm{~and~}     \begin{pmatrix}
        I_{3^{\prime}} \\
        I_{2^{\prime}} \\
        I_{1^{\prime}} \\
    \end{pmatrix}
    = \bar{S}
    \begin{pmatrix}
        O_{3^{\prime}} \\
        O_{2^{\prime}} \\
        O_{1^{\prime}} 
    \end{pmatrix}.
    \label{eqn:scat1}
\end{equation}
All other equations from the uniform case describing the phases acquired as the waves move through the unit cell apply identically here. 

The $4 \times 4$ scattering matrix has the same block matrix structure as the uniform case (Eq.~\ref{eq:sigma}), but with a different functional form for the matrix elements, and the periodic boundary conditions we impose are also the same. We therefore solve the same equation (Eq.~\ref{eq:main}), but now including the new $\Sigma$ matrix elements. We analytically solve the equation as a function of $\zeta$, with the dispersive bands for the example $\zeta = 3/2$ case shown in Fig.~\ref{fig.3}(a) and (b). Generally, for $\zeta \neq 1$, we find that the two Dirac points are displaced around the Brillouin zone and remain gapless (Fig.~\ref{fig.3}(c) and (d)). However, as $\zeta$ is decreased below some critical value, the Dirac points merge in pairs and annihilate, opening a gap between the dispersive bands. This behaviour is qualitatively similar to that described in Ref.~\onlinecite{Hou2015}, suggesting that a similar hidden symmetry may be responsible for protecting the Dirac points under lattice strain. The flat band remains unperturbed as strain is introduced, and it still meets the dispersive bands at the origin.

These results are straightforwardly connected to the theory of artificial magnetic fields in strained honeycomb lattices~\cite{Vozmediano2010}: the $\mathbf{k}$-space displacement of the Dirac points can in fact be interpreted as arising from a vector potential $\mathbf{A}$. Given the symmetry of our system under time-reversal, the displacement is opposite on the two Dirac points, which corresponds to a valley-dependent vector potential. A magnetic field is then naturally obtained by means of a spatially-dependent impedance modulation, that provides a spatially-dependent $\mathbf{A}(\mathbf{r})$. This opens the possibility of exploring magnetic models and, for instance, observing the peculiar square-root spectrum of the relativistic Landau levels~\cite{Salerno2015}.

\section{Engineering isolated flat bands}
\label{sec:lengths}

As a change of impedance is not enough to energetically isolate the flat bands, we now investigate the effect of our other main control knob, namely the cable lengths. In particular, we consider a lattice with two different lengths of cable, both with the same impedance. The cables running between the two connectors in each unit cell (the central cable) now have length $l_2$, and the other cables have length $l_1$ (Fig.~\ref{fig.1}). The general approach to the calculation of the Bloch wave dispersion is the same as the other two cases, but the equations describing the phases that the modes acquire as they propagate in the cable segments now depend on the two different lengths. 

The $4 \times 4$ unit cell scattering matrix, again, has the same block structure as the uniform case but with matrix elements that now depend on the ratio of lengths $\lambda \equiv l_1 / l_2$. The periodic boundary conditions we then apply are similar to previous cases, but including the two different cable lengths:
\begin{equation}
   \begin{pmatrix}
       I_{1x} \\
       O_{1x}
   \end{pmatrix} 
   =
   e^{ik_x(l_1 + l_2)}
   \begin{pmatrix}
       O_{0x} \\
       I_{0x}
   \end{pmatrix},
\end{equation}
and:
\begin{equation}
   \begin{pmatrix}
       I_{1y} \\
       O_{1y}
   \end{pmatrix} 
   =
   e^{i(k_xl_2 + k_yl_1)}
   \begin{pmatrix}
       O_{0y} \\
       I_{0y}
   \end{pmatrix},
\end{equation}
We therefore, again, solve the same fundamental equation as the other cases. We focus on the analytical solution to the next-simplest case relative to the solved $\lambda = 1$ situation, namely $\lambda = 2$. This equation produces flat bands with $e^{-i\omega l_2/2c} = \pm 1$, and dispersive bands whose dispersion is the solution to:
\begin{multline}
    9\cos(2\omega l_2/c) + 8\cos(\omega l_2/c) +\\ - 8\cos\left(\frac{\omega l_2}{2c}\right)\left(\cos\left(\frac{3k_xl_2}{2}\right)+\right.\\ + \left.\cos\left(\frac{k_xl_2}{2} + k_yl_2\right)\right)+ \\ + 3 - 4\cos(k_xl_2 - k_yl_2) = 0.
\end{multline}
The first equation yields flat bands for $\omega l_2/c = 2\pi n$ for $n = 1,2,3,...$, and the second equation can be solved analytically to produce four distinct dispersive bands in the interval $\omega_{PV} l_2/c \in [-\pi, \pi]$, which are then repeated as $\omega l_2/c = \omega_{PV} l_2/c + 4\pi n$, with $n = 0, 1, 2, ...$. 

The overall interleaved structure of the bands is shown in the cut in Fig.~\ref{fig.4}(a), and Fig.~\ref{fig.4}(b) - (e) show the structure of the dispersive bands in detail. Crucially, we find a family of energetically-isolated flat bands with $\omega l_2/c= 2\pi, 6\pi, 10\pi, ...$. Since any excitations from these bands will have no kinetic energy, any interaction effects, which could be introduced via the use of nonlinear circuit elements, will dominate the physics. This opens the exciting prospect of studying interacting models using this platform, possibly in combination with magnetic field effects that could be engineered by control of the cable impedance.

\section{Conclusions}

In this work, we proposed using lattices formed from standard, commercially-available coaxial cables and connector elements to engineer photonic lattice models. As an illustrative example, we calculated the band structure for a brick wall lattice of cables and T-connectors. 

In agreement with the Kagome geometry of the effective photonic lattice corresponding to this physical lattice, we showed that this model hosts one flat band, and two dispersive bands that meet at two Dirac points in each Brillouin zone. This set of three principal bands is then repeated periodically in frequency. We then showed how these features can be controlled by modulating the cable impedances and lengths. In particular, we can displace the Dirac points around the Brillouin zone, which allows us to engineer artificial magnetic fields. We can also isolate an infinite family of flat bands, which suggests exciting prospects when nonlinearities are introduced.

Our results only scratch the surface of the proposed experimental platform, and a number of exciting future research directions are readily apparent. In the short-term, our mathematical formalism should be extended to allow for the analysis of symmetries and their effect on band crossings. In a typical lattice model, this would be approached by finding unitary symmetry operators that commute with the Hamiltonian, but the equivalent of this in our formalism without an explicit Hamiltonian is an open problem. Extending the formalism in this way would allow us to assess the robustness of the band-crossing at the origin and of the Dirac points when the lattice is distorted.

In the long-term, two major future directions are immediately suggested by our current results: firstly, one can exploit the artificial magnetic field to realise magnetic models displaying, for instance, a sequence of quantized Landau levels with the peculiar square-root energy spacing typical of relativistic models. 
Secondly, nonlinear circuit elements could be introduced to investigate interaction effects, with a particular emphasis on the flat band case, where interaction effects should dominate. Going further, gain elements could also be included, with a view towards non-Hermitian effects such as topological lasing. Overall, our results readily suggest a variety of interesting future directions, with a view towards unlocking the full potential of this new quantum simulation platform.

\acknowledgments

We would like to thank Alicia Koll\'ar for useful and engaging discussions. I.C. acknowledges financial support from the Provincia Autonoma di Trento, from the Q@TN Initiative, and from the PNRR MUR project PE0000023-NQSTI. C.O. and H.M.P. are supported by the Royal Society via grants UF160112, RGF/EA/180121 and RGF/R1/180071, and by the Engineering and Physical Sciences Research Council [grant number EP/W016141/1].

\end{document}